\RequirePackage[2020-02-02]{latexrelease}
\documentclass[sn-mathphysaps,floatfix]{revtex4}
\usepackage{graphicx}
\usepackage{dcolumn}
\usepackage{bm}
\usepackage{float}
\usepackage{placeins}
\usepackage{amsmath}
\usepackage{romannum}
\usepackage{csquotes}
\usepackage{tikz}
\usetikzlibrary{shapes}
\usepackage{hyperref}
\hypersetup{
    colorlinks = true,
    citecolor = blue,
    linkcolor = blue,
     urlcolor = blue,
}
\newcommand{\markerone}{\raisebox{0.5pt}{\tikz{\node[star, star points=5, star point ratio=2.25, draw, inner sep=0.15em, anchor=outer point 3,scale=0.6,fill=green](){};}}}
\newcommand{\markertwo}{\raisebox{0.6pt}{\tikz{\node[draw,scale=0.5,rectangle,fill=blue](){};}}}
\newcommand{\markerthree}{\raisebox{0.5pt}{\tikz{\node[draw,scale=0.5,circle,fill=red](){};}}}        
\begin{document}
\title{A Strongly Correlated Quantum-Dot Heat Engine with Optimal Performance: An Non-equilibrium Green's function Approach}
\author{\bf{Sachin Verma$^1$}}
\author{\bf{Ajay$^1$}}
\affiliation{$^1$Department of Physics, Indian Institute of Technology, Roorkee, Uttarakhand, 247667, India\\
\text{Email:} sverma2@ph.iitr.ac.in and ajay@ph.iitr.ac.in}
\begin{abstract}
We present an analytical study of a strongly correlated quantum dot-based thermoelectric particle-exchange heat engine for both finite and infinite on-dot Coulomb interaction. Employing Keldysh's non-equilibrium Green's function formalism for different decoupling schemes in the equation of motion, we have analyzed the thermoelectric properties within the non-linear transport regime. As the simplest mean-field approximation is insufficient for analyzing thermoelectric properties in the Coulomb blockade regime, one needs to employ a higher-order approximation to study strongly correlated QD-based heat engines. Therefore initially, we have used the Hubbard-\Romannum{1} approximation to study the quantum dot level position ($\epsilon_d$), thermal gradient ($\Delta T$), and on-dot Coulomb interaction ($U$) dependence of the thermoelectric properties. Furthermore, as a natural extension, we have used an approximation beyond Hubbard-\Romannum{1} in the infinite-$U$ limit (strong on-dot Coulomb repulsion) to provide additional insight into the operation of a more practical quantum dot heat engine. Within this infinite-$U$ limit, we examine the role of the symmetric dot-reservoir tunneling ($\Gamma$) and external serial load resistance ($R$) in optimizing the performance of the strongly correlated quantum dot heat engine. Our infinite-$U$ results show a good quantitative agreement with recent experimental data for a quantum dot coupled to two metallic reservoirs.
\end{abstract}

\maketitle
\pagenumbering{arabic}
 
\section{Introduction}
\label{sec: first}
\begin{figure*}[!htb]
\centering
\includegraphics
  [width=0.65\hsize]
  {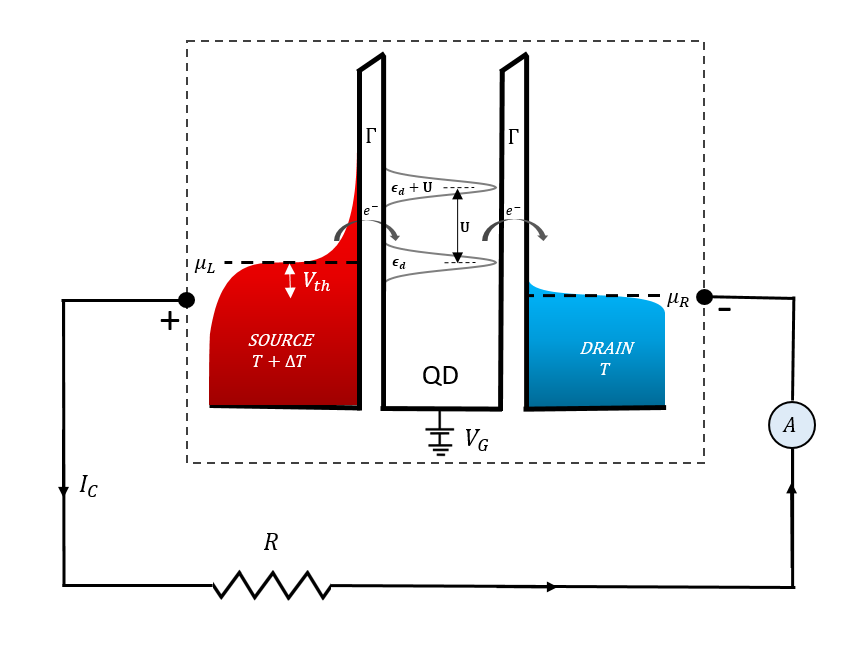}
\caption {Schematic circuit diagram for a hybrid normal-metal quantum-dot particle exchange heat engine. The gate tunable discrete energy levels of QDs serve as ideal energy filters for electron transport, resulting in high thermoelectric performance. Due to a finite temperature difference $\Delta{T}$, electrons tunnel from the left hot reservoir to the QD energy level and then to the right cold reservoir producing a thermovoltage $V_{th}$. Due to this thermovoltage the power generated by the heat engine is available for consumption in $R$, which also includes the resistance of other external circuit elements.}
\label{fig:1}
\end{figure*}

The heat and thermoelectric transport properties of hybrid quantum dot(s) nano-structures are recently attracting great attention experimentally and theoretically due to their potential applications in solid-state thermal devices\cite{DiSalvo1999,Giazotto2006,Dubi2011,Radousky2012,Haupt2013,Sothmann2014,Pekola2021}. One such hybrid nanostructure is a quantum dot (QD) particle-exchange heat engine. These particle-exchange heat engines consist of quantum dot(s) connected to metallic source and drain reservoirs by tunnel junctions\cite{Beenakker1992,Andreev2001,Humphrey2002,Turek2002,Koch2004,Humphrey2005,Nakpathomkun2010,Mani2011}. Due to quantum confinement and Coulomb blockade (CB) effects on QD, these low-dimensional heat engines are more efficient at converting thermal energy into electrical energy than their bulk counterparts\cite{Hicks1993-96,Mahan1996,Broido1997,Chen2003,Mao2016,Majidi2022}. The enhancement of thermoelectric efficiency in QD heat engines is due to the strong violation of the Wiedemann-Franz law\cite{Boese2001,Dong2002,Krawiec2006,Zianni2007,Kubala2008,Murphy2008,Franco2008} and the significant drop in lattice thermal conductivity\cite{Balandin1998,Khitun2000,Liu2009,Tsaousidou2010}.\\
In QD heat engines, electrons tunnel from the left hot reservoir to the QD and then to the right cold reservoir, resulting in a thermovoltage, or voltage caused by the temperature difference between the reservoirs due to the Seebeck effect. Thus, the temperature difference or thermal gradient can drive an electric current in the external circuit. The power output generated by the heat engine is available for consumption in an external serial load $R$ [Fig.\hyperref[fig:1]{1}]. This thermoelectric power conversion could be useful for future energy harvesting in nano-electronic devices and quantum technologies\cite{Radousky2012,Sothmann2014}. Apart from their potential practical uses, these hybrid QD systems can also be used as a testing ground for various theoretical techniques to further analyse the transport properties of novel strongly correlated materials.\\
Thermovoltage, thermopower and thermoelectric performance of hybrid normal-metal quantum-dot (N-QD-N) heat engines have been widely studied using various theoretical techniques such as equation of motion\cite{Liu2010,Swirkowicz2009,Wierzbicki2010,Karlstrom2011,Yan2014,Sierra2014,Zimbovskaya2015,Kuo2017}, slave Boson\cite{Krawiec2006,Krawiec2007}, perturbation theory\cite{Boese2001,Dong2002,Kubala2006,Kubala2008}, master equation approach\cite{Beenakker1992,Zianni2007,Zianni2008,Zianni2010,Sanchez2011,Muralidharan2012,Rossello2017,Erdman2017}, Wilson's numerical renormalization-group (NRG)\cite{Costi1994,Costi2010,Andergassen2011} and functional renormalization-group (FRG) method\cite{Kennes2013}. The majority of these papers focused on the linear response regime, i.e., for small thermal gradients ($\delta T\rightarrow0$) and small voltage biases ($eV\rightarrow0$). In the linear response regime, the thermoelectric conversion efficiency is characterized by a dimensionless figure of merit $ZT=S^2GT/K$, where $T$ is the absolute temperature, $S$ is the Seebeck coefficient (or thermopower), $G$ is electrical conductance, and $K$ is thermal conductance(includes both electron and phonon contributions). For larger $ZT$, the thermoelectric conversion efficiency ($\eta$) approaches its upper bound, i.e., the Carnot efficiency ($\eta_c$). The efficiency for short-lived quantum states ($\Gamma/k_BT>>1$, where $k_B$ is Boltzmann constant) is relatively low, while it approaches the Carnot efficiency for long-lived quantum states (i.e., for $\Gamma/k_BT<<1$). The enhancement in efficiency is caused by sequential tunneling events (i.e., lowest order incoherent tunneling processes), which dominate transport at relatively higher temperatures (but $U>>k_BT$) and violate the Wiedemann-Franz law by suppressing thermal conductance in comparison to charge conductance.\\
The influence of asymmetric dot-reservoir coupling\cite{Krawiec2007}, electron-electron interactions\cite{Zianni2007,Zianni2008,Kubala2008,Murphy2008,Liu2010,Yan2014,Zimbovskaya2015,Taylor2015}, Kondo effect\cite{Boese2001,Dong2002,Franco2008,Costi2010,Andergassen2011}, energy spectrum of the QD\cite{Zianni2008,Kuo2017,Erdman2017}, electron-phonon interactions\cite{Zianni2010,Ren2012,De2016}, and the Fano effect\cite{Wierzbicki2011,Bevilacqua2016,Taniguchi2020} on the thermoelectric properties of QD heat engines has also been investigated in the linear and non-linear regime. Krawiec et al.\cite{Krawiec2007} studied the influence of asymmetric dot-reservoir coupling ($\Gamma_L\neq\Gamma_R$) on thermoelectric properties and showed that in the linear regime, the thermopower $S$ and figure of merit $ZT$ are unaffected by asymmetry. On the other hand, the non-linear $S$ strongly depends on asymmetry. The effects of electron-electron Coulomb interaction on thermoelectric properties were analyzed in several works. However, in the non-linear regime, the effect of Coulomb interaction on the thermovoltage or thermopower, thermocurrent, and thermoelectric power has been studied in a few works\cite{Yan2014,Sierra2014,Zimbovskaya2015}. At very low temperatures ($T\leq T_K$, where $T_K$ is the Kondo temperature), where the Kondo effects develop and the ground state shows a Fermi liquid behavior, the Wiedemann-Franz law is recovered\cite{Dong2002,Franco2008,Costi2010}. Both $S$ and $ZT$ for $T\leq T_K$ are rather small and can be enhanced in the relatively high-temperature regime ($T>>T_K$). The effect of multi-levels of QD on the thermoelectric properties has been investigated in Ref. \cite{Zianni2008,Kuo2017,Erdman2017} within the sequential tunneling regime. Also, at finite electron-phonon coupling, the decrease in the power factor $S^2GT$ and increase in $K$ can significantly reduce $ZT$\cite{Zianni2010}. However, the electron-phonon coupling is small at low temperatures and can be effectively suppressed in the experiments. Recently, Taniguchi\cite{Taniguchi2020} showed that in the non-linear regime, even if the temperature is much lower than the dot-reservoir tunneling rate ($\Gamma/k_BT>>1$), which is unfavorable for good thermoelectric conversion efficiency, one can still achieve reasonably good thermoelectric performance by regulating quantum coherence via the Fano effect. Other theoretical studies have also shown that optimal thermoelectric operation is possible in the non-linear transport regime\cite{Esposito2009,Leijnse2010,Wierzbicki2012,Whitney2014,Talbo2017,Sanchez2016}. On the other hand, the experimental studies on thermoelectric transport properties of QD heat engine is limited\cite{Svilans2016,Staring1993,Dzurak1997,Appleyard2000,Small2003,Scheibner2005,Scheibner2007,Svensson2012,Svensson2013,Josefsson2018,Jaliel2019,Dorsch2021,Svilans2018,Erdman2019}. Only a couple of these experiments have focused on measuring power output and corresponding thermoelectric efficiencye\cite{Josefsson2018,Jaliel2019,Dorsch2021}. Recently, Josefsson et.al.\cite{Josefsson2018} demonstrated that for $\Gamma<<k_BT$ with strong on-dot Coulomb interaction($U>>k_BT$), a QD heat engine can achieve a thermoelectric conversion efficiency close to 70\% of the Carnot limit. These experiments also required additional theoretical research into the practical optimization of QD-based heat engines, particularly the impact of non-linear effects and external circuit elements on the performance of these nano-devices. \\
In the present work, we contribute to previous studies by analyzing the non-linear transport regime as well as the effect of the external load resistance on the practical optimization of the performance of QD-based heat engines. We consider a single-level Coulomb blockade quantum dot connected to the normal metallic source and drain reservoirs [see Fig.\hyperref[fig:1]{1}]. By using non-equilibrium Keldysh formalism within various decoupling schemes in the Green's function equation of motion technique (Hartree-Fock mean-field, Hubbard-\Romannum{1} and beyond Hubbard-\Romannum{1} for infinite-$U$ limit), we have analyzed the thermoelectric transport properties within the linear and non-linear transport regime. Furthermore, we limit our attention to cases where $T>>T_K$, i.e., within the non-Kondo regime. The primary aim of the present study is to understand the effect of Coulomb interaction, applied thermal gradient, dot-reservoir tunneling rate, and external load resistance on the thermoelectric transport properties (especially on the power output and corresponding efficiency) in the non-linear regime. We pay particular attention to analyzing the effect of the external load $R$ on the optimization of the power output and corresponding thermoelectric efficiency and reproducing the experimental results reported in Ref.\cite{Josefsson2018}.\\
The rest of the paper is organized as follows. The preceding section, \hyperref[sec: second]{\Romannum{2}}, contains the model Hamiltonian and theoretical description. In section-\hyperref[sec: third]{\Romannum{3}}, the numerical results and discussion of the linear and non-linear transport regimes are provided. In section-\hyperref[sec: fourth]{\Romannum{4}} we conclude the present work.
\section{Model Hamiltonian and theoretical description}
\label{sec: second}
The system under consideration consists of a single-level quantum dot coupled between normal metallic source and drain reservoirs. Such system is modelled by single impurity Anderson Hamiltonian in second quantization formalism,
\begin{equation} \label{eq:1}
\hat{H}=\hat{H}_{reservoirs}+\hat{H}_{QD}+\hat{H}_{tunnel} 
\end{equation}
where,\\
$\hat{H}_{reservoirs}=\sum_{k\sigma,\alpha\in L,R}(\epsilon_{k,\alpha}\hat{c}^\dagger_{k\sigma,\alpha}\hat{c}_{k\sigma,\alpha})$ represents the  Hamiltonian of the metallic reservoirs, where $\epsilon_{k,\alpha}$ is the kinetic energy of the electrons in the reservoirs, and $c_{k\sigma}(c^\dagger_{k\sigma})$ is the annihilation(creation) operator of an electron with spin $\sigma$ and wave vector $\vec{k}$.\\
$\hat{H}_{QD}=\sum_{\sigma} \epsilon_{d}\hat{n}_{d\sigma}+U\hat{n}_{d\uparrow}\hat{n}_{d\downarrow}$  represents the Hamiltonian of the quantum dot with $n_{d\sigma}=d_\sigma^\dagger d_\sigma$ as the number operator and  $d_\sigma(d^\dagger_\sigma)$ is the annihilation(creation) operator of electron with spin $\sigma$. The quantum dot consists of a single electronic level of energy $\epsilon_d$ and can be occupied by up to two electrons, i.e., $\sigma\in\uparrow,\downarrow$. We also consider the intradot or on-dot electron-electron Coulomb repulsion with the interaction strength $U$.\\
$\hat{H}_{tunnel}=\sum_{k\sigma,\alpha}(V_{k,\alpha}\hat{d}^\dagger_{\sigma}\hat{c}_{k\sigma,\alpha}+{V^\ast_{k,\alpha}}\hat{c}^\dagger_{k\sigma,\alpha}\hat{d}_{\sigma})$ represents the tunneling Hamiltonian between the QD and reservoirs with coupling strength $V_{k,\alpha}$.\\
Using the non-equilibrium Green's function formalism, the electric (charge) current $I_C$ and heat current $J_Q$ from left to right reservoir across the QD for symmetric tunnelling rate ($\Gamma_L=\Gamma_R=\Gamma$) can be expressed as \cite{Dubi2011,Pekola2021,Meir1992,Haug2008}
\begin{equation} \label{eq:2}
I_C = \frac{2e}{h} \int{\left[f_L(\omega-\mu_L)-f_R(\omega-\mu_R)\right] T(\omega)d\omega}
\end{equation}
 \begin{equation} \label{eq:3}
J_Q = \frac{2}{h} \int{(\omega-\mu_L)\left[f_L(\omega-\mu_L)-f_R(\omega-\mu_R)\right] T(\omega)d\omega}
\end{equation}
where $T(\omega)=\Gamma^2 |G_{11\sigma}^r(\omega)|^2$ is the tunnelling amplitude or transmission function and $f_{\alpha}(\omega\mp\mu_{\alpha})=\left[{exp((\omega\mp\mu_{\alpha})/k_BT_{\alpha})+1}\right]^{-1}$ is the Fermi-Dirac distribution function of $\alpha$-reservoir with chemical potential $\pm\mu_{\alpha}$ and temperature $T_{\alpha}$. Also, $G_{11\sigma}^r(\omega)=\langle\langle{d_{\sigma}|d^\dagger_{\sigma}}\rangle\rangle$ is the Fourier transformation of the single particle retarded Green's function for the QD in time domain (i.e. $G_{11\sigma}^r(t-t^{'})=\mp i\theta(t-t^{'})\langle{\{d_{\sigma}(t),d_{\sigma}^{\dagger}(t^{'})\}}\rangle$). The derivation for the expression of the single particle retarded Green's function within Hartree-Fock mean-field (Eq.\hyperref[eq:A3]{(A3)}), Hubbard-\Romannum{1} (Eq.\hyperref[eq:A5]{(A5)}) and beyond Hubbard-\Romannum{1} (Eq.\hyperref[eq:B7]{(B7)}) decouplings in the equation of motions (EOMs) are given in Appendix~\hyperref[Appendix: A]{A} and \hyperref[Appendix: B]{B}.\\
Once the single particle retarded Green's function of QD is known the average occupation on the QD ($\langle{n_{\sigma}}\rangle$=$\langle{n_{\bar{\sigma}}}\rangle$ for non-magnetic system) is calculated by using the self-consistent integral equation of the form
\begin{equation} \label{eq:4}
\langle{n_{\sigma}}\rangle=\frac{-i}{2\pi}\int^{\infty}_{-\infty}G^{<}_{11\sigma}(\omega) d\omega
\end{equation}
where ${G}^{<}_{11\sigma}(\omega)$ is the Fourier transformation of the single particle lesser Green's function for the QD in time domain (i.e. $G^{<}_{11\sigma}(t-t^{'})=i\theta(t-t^{'})\langle{d_{\sigma}^{\dagger}(t^{'}),d_{\sigma}(t)}\rangle$) and obeys the Keldysh equation in matrix form\cite{Haug2008,Keldysh1965},
 \begin{equation} \label{eq:5}
{\bf{G}}^{<}_{\sigma}(\omega)={\bf{G}}^{r}_{\sigma}(\omega){\bf{\Sigma}}^{<}_{\sigma}(\omega){\bf{G}}^{a}_{\sigma}(\omega)
\end{equation}
where $\bf{G}^{r}_{\sigma}(\omega)$ is retarded Green's function matrix, ${\bf{G}}^{a}_{\sigma}(\omega)=\left[{\bf{G}}^{r}_{\sigma}(\omega)\right]^{\dagger}$ is the advanced Green's function matrix and ${\bf{\Sigma}}^{<}_{\sigma}(\omega)= -\sum_{\alpha}\left[{\bf{\Sigma}}^{r}_{\alpha}-{\bf{\Sigma}}^{a}_{\alpha}\right]{\bf{f}}_{\alpha}(\omega)$ is the lesser self energy matrix.
Thus,
 \begin{equation} \label{eq:6}
 {\bf{G}}^{<}_{\sigma}(\omega)=
  \begin{pmatrix}
 G_{11\sigma}^{r}(\omega) & 0 \\
 \\
 0 & G_{22\sigma}^{r}(\omega) \\
 \end{pmatrix}
  \begin{pmatrix}
\sum_{\alpha} \frac{i\Gamma_{\alpha}}{2} f_L(\omega-\mu_{\alpha}) & 0 \\
 0 & \sum_{\alpha} \frac{i\Gamma_{\alpha}}{2} f_L(\omega+\mu_{\alpha}) \\
 \end{pmatrix}
 \\
  \begin{pmatrix}
 G_{11\sigma}^{a}(\omega) & 0 \\
 \\
 0 & G_{22\sigma}^{a}(\omega) \\
 \end{pmatrix}
\end{equation}
Thus, the lesser Green's function for single particles/electrons on the quantum dot is given by,
 \begin{equation} \label{eq:7}
G^{<}_{11\sigma}(\omega)=\frac{i}{2}\left[\Gamma_L f_L(\omega-\mu_L)+\Gamma_R f_R(\omega-\mu_R)\right] |G_{11\sigma}^{r}(\omega)|^2
\end{equation}
For the QD based particle-exchange heat engine the temperature of the left reservoir is considered as  $T_L=T+\Delta T$ while the right reservoir remains at the background temperature $T_R=T$. Due to this temperature difference electrons moves from left hot reservoir to the right cold reservoir and creates a potential difference ($V_{th}=(\mu_L-\mu_R)/e$) due to accumulation of electron on the right reservoir and positive charge to the left reservoir [see Fig.\hyperref[fig:1]{1}]. Thus a thermally induced charge current flow through the external circuit. Further, to achieve optimal thermoelectric conversion we consider symmetric dot-reservoir tunneling rate i.e., $\Gamma_L=\Gamma_R=\Gamma$\cite{Leijnse2010}.\\
Usually, the thermovoltage ($V_{th}$) is determined from the open circuit condition $I_C(V_{th},\Delta T)=0$ i.e. a external reverse bias voltage $V_{ext}=(\mu_R-\mu_L)/e$ is applied to counteract the thermally induced charge current. The thermopower $S={V_{th}}/{\Delta T}$ and thermal conductance $K = {J_Q}/{\Delta T}$ are then calculated. In open circuit condition, the maximal power generated by the heat engine is $P_{max} = -(I_C)_{V_{max}}V_{max}=-(I_C)_{V_{th}/2}(V_{th}/2)$. However, if the external voltage source is replaced by the load resistance ($R$) then the current has to self-consistently satisfy the equation,
\begin{equation} \label{eq:8}
I_C(V_{th},\Delta T) + V_{th}/R = 0
\end{equation}
Eq.\hyperref[eq:8]{(8)} is solved numerically to obtain thermovoltage $V_{th}$.\\
The finite power output $P=-I_CV_{th}=I_C^2R$ generated by the QD heat engine dissipates across the load resistance $R$. Thus only optimizing $V_{th}$ and $\epsilon_d$ are not sufficient to reach the maximum power because $P$ also depends on external load resistance $R$. The efficiency of heat engine at maximal power output is calculated by using  $\eta_{P_{max}} = P_{max}/J_Q$, which can be normalized by the Carnot efficiency $\eta_C = 1-T_R/T_L$. It is also important to note that we only consider thermal contribution by electrons and neglected the phonon contribution which is small at low temperatures and can also be suppressed effectively in hybrid QD devices.
\section{Result and discussion}\label{sec: third}
The equations derived in the previous section within non-linear transport regimes are numerically solved using MATLAB, with $\Gamma_0$ (in $meV$) as the energy unit.
\subsection{Linear response transport regime}
The thermoelectric transport properties of a QD-based particle-exchange heat engine have been extensively studied in the linear response regime, as discussed in section-\hyperref[sec: first]{\Romannum{1}}. In this subsection, we plot the thermoelectric quantities using linear response relations\cite{Mahan2000} to compare the Hartree-Fock (HF) mean-field and Hubbard-\Romannum{1} approximations which are used to analyze the Coulomb blockade effect.\\
\begin{figure*}[!htb]
\centering
\includegraphics
  [width=0.85\hsize]
  {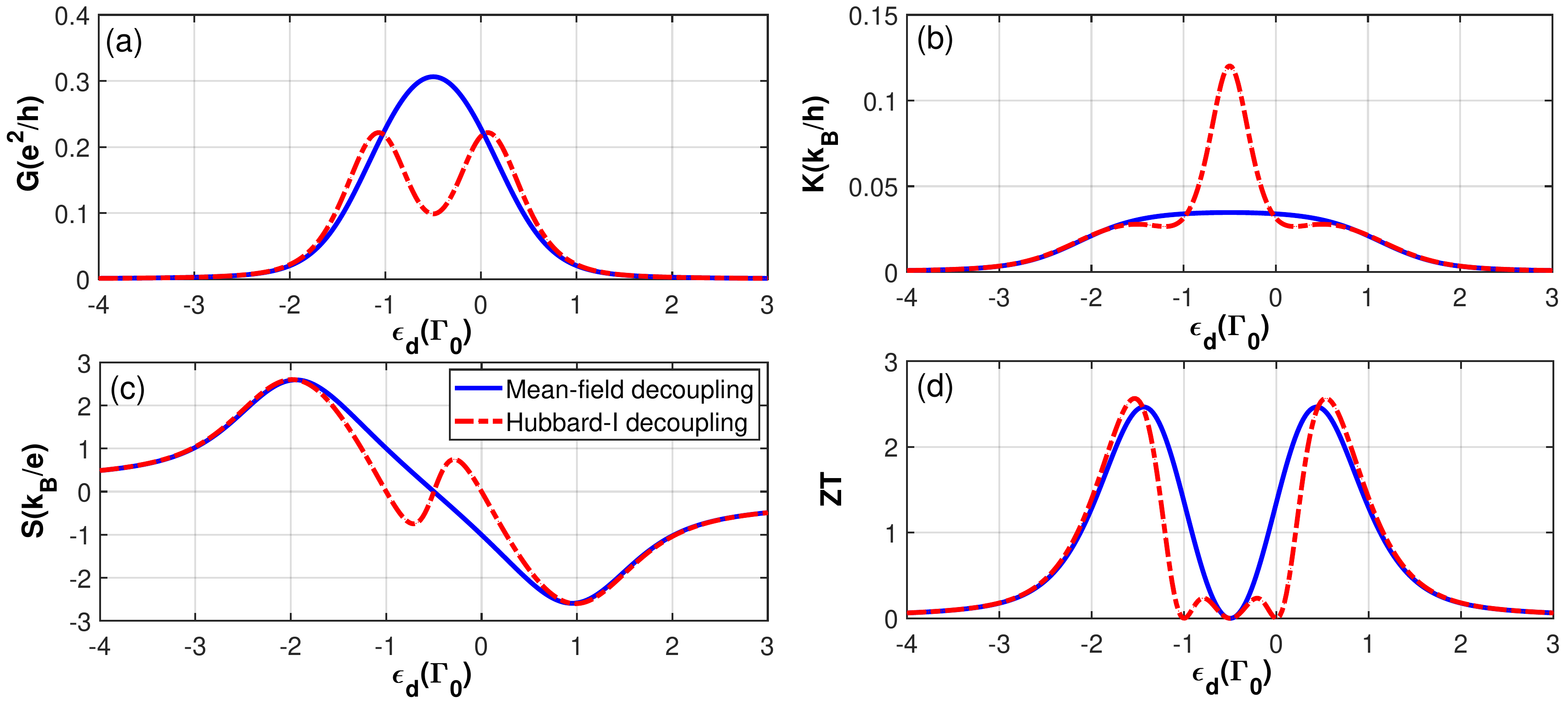}
\caption {Thermoelectric transport quantities (electrical conductance $G$, thermal conductance $K$, thermopower $S$, and figure of merit $ZT$) of a N-QD-N particle exchange heat engine as a function of QD energy level $\epsilon_d$ within HF mean-field and Hubbard-\Romannum{1} decoupling schemes in EOMs. The parameters are $U=1.0\Gamma_0$, $\Gamma=0.1\Gamma_0$ and $k_BT=0.2\Gamma_0$.}
\label{fig:2}
\end{figure*}
Fig.\hyperref[fig:2]{2} shows the plots for the linear thermoelectric quantities as a function of QD energy level ($\epsilon_d$) in the Coulomb blockade regime using HF mean-field and Hubbard-\Romannum{1}  decoupling techniques in EOMs (see Appendix~\hyperref[Appendix: A]{A}). These plots show that the HF mean-field approximation only reproduces the gross features of the thermoelectric quantities by simply shifting the non-interacting resonant QD energy level from $\epsilon_d$ to $\epsilon_d+U\langle{n_{\sigma}}\rangle$. These HF results deviates from Hubbard-\Romannum{1} results for $-1.5\Gamma_0\leq\epsilon_d\leq0.5\Gamma_0$. Thus, the HF mean-field approximation, previously applied to study the spectral and charge transport properties at absolute zero temperature in the weak Coulomb blockade regime, becomes insufficient for analyzing finite temperature thermoelectric transport properties. The Hubbard-I decoupling approach, on the other hand, takes into account higher-order tunneling processes due to electron-electron interaction on the QD  and consists of two effective energy levels, one at $\epsilon_d$  and the second at $\epsilon_d+U$. Thus, one can get characteristic features in the thermoelectric properties by tuning these effective levels (by applying gate voltage) above and below the Fermi energy. For example, electrical conductance $G$ shows two broad peaks at the resonance energies (i.e., when one of the effective levels lies at the Fermi energy), and thermopower $S$ changes sign three times, i.e., at resonance energies, and the electron-hole symmetry point ($\epsilon_d=-U/2$). Therefore in the present study, we have used the Hubbard-\Romannum{1} approximation to investigate the effect of finite on-dot Coulomb repulsion on the thermoelectric transport properties of the QD heat engine.
\subsection{Non-linear transport regime}
In this subsection, we study the effect of (1) thermal gradient $\Delta{T}$, on-dot Coulomb interaction $U$, and (2) external load resistance $R$ on the thermoelectric transport properties of a QD heat engine beyond the linear response regime. We have used both Hubbard-\Romannum{1} (for finite-$U$ limit) and beyond Hubbard-\Romannum{1} approximation (for infinite-$U$ limit). Finally, the theoretical results for a QD heat engine connected with external serial load resistance are compared with the recent experimental data.
\subsubsection{Effect of thermal gradient and Coulomb interaction}
\begin{figure*}[!htb]
\centering
\includegraphics [width=0.85\hsize]{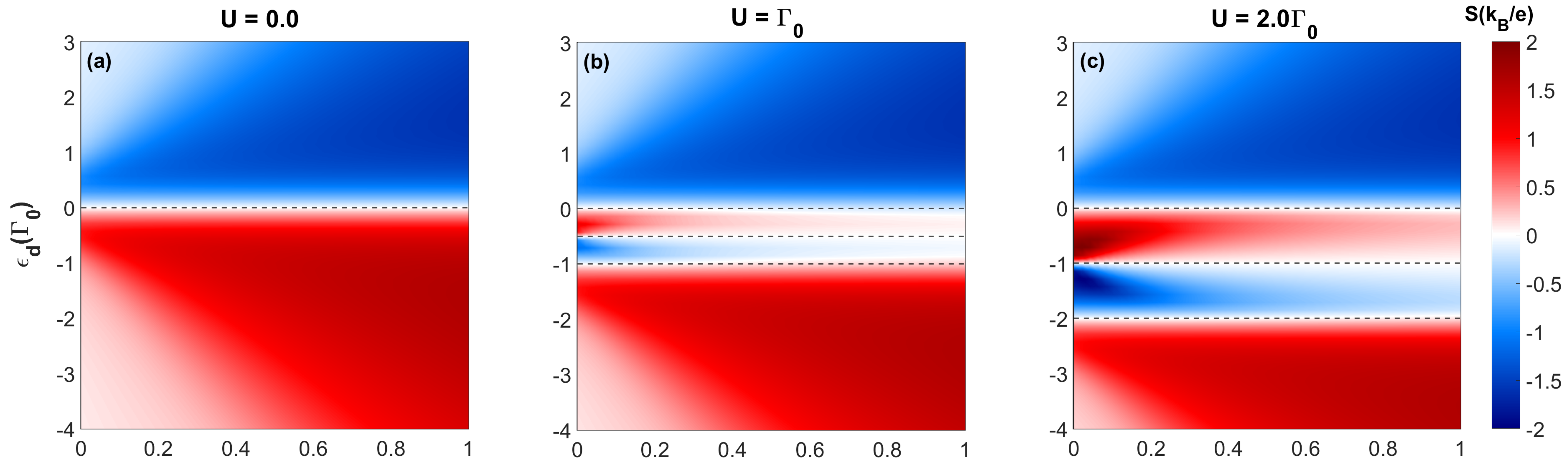}
\includegraphics [width=0.85\hsize]{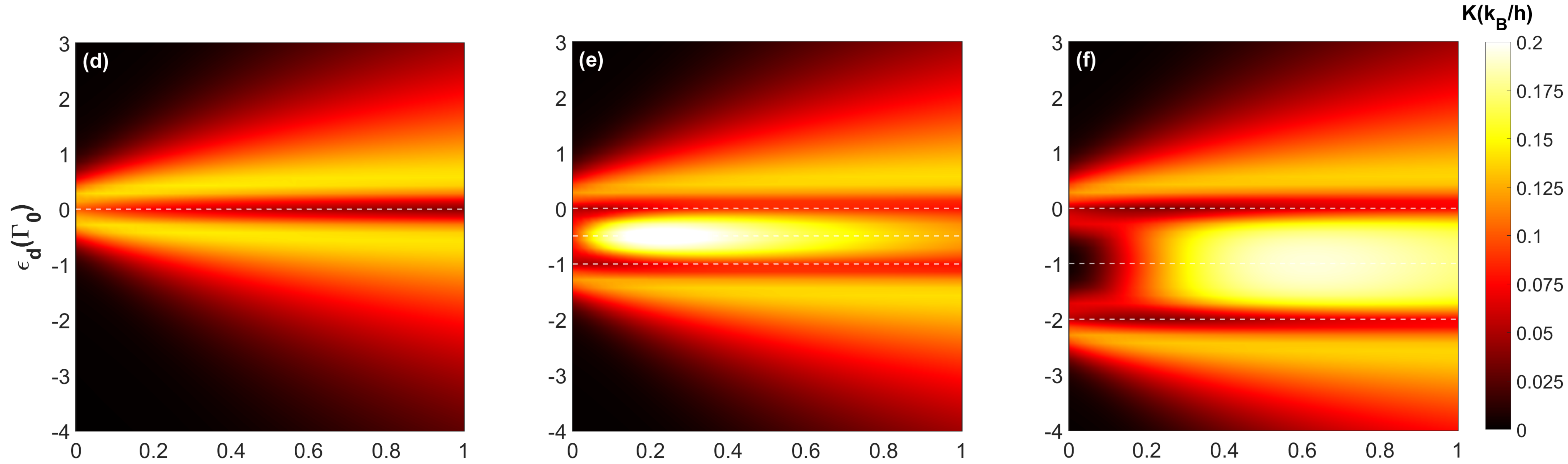}
\includegraphics [width=0.85\hsize]{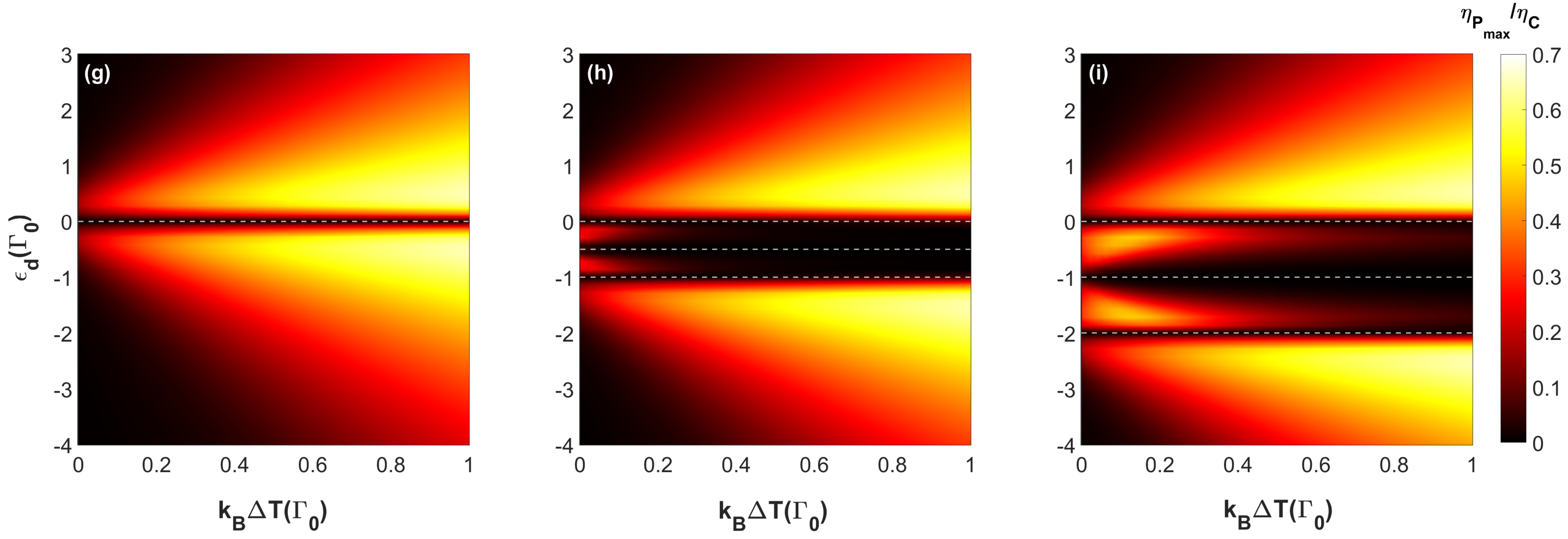}
\caption {Variation of non-linear (a)-(c) thermopower($S$), (d)-(f) thermal conductance($K$), and (g)-(i) normalized efficiency at maximum power output($\eta_{P_{max}}$) as a function of temperature difference ($\Delta T$) between the reservoirs and the QD energy level ($\epsilon_d$) for different on-dot Coulomb repulsion $U$ with $k_BT= 0.2\Gamma_0$, $\Gamma=0.1\Gamma_0$, and $\Gamma_0$ is the energy unit. The results are obtained within the Hubbard-\Romannum{1} approximation for finite $U$ by using Green's function EOM technique.}
\label{fig:3}
\end{figure*}
In Fig.\hyperref[fig:3]{3}, we have shown the non-linear thermopower $S$, electronic thermal conductance $K$ and thermoelectric efficiency at maximum power output $\eta_{P_{max}}$ as a function of QD energy level $\epsilon_d$ and applied thermal gradient $\Delta{T}$ for different on-dot Coulomb interaction $U$. Here, we have used the open circuit condition, i.e., $I_C(V_{th},\Delta T)=0$ to calculate the thermoelectric properties of the QD heat engine.\\
The thermopower in fig.\hyperref[fig:3]{3}(a-c) is antisymmetric around $\epsilon_d=-U/2$ due to electron-hole symmetry, and positive (negative) thermopower indicates that holes (electrons) are the majority charge carriers. Due to finite on-dot Coulomb repulsion $U$, two effective levels lie at $\epsilon_d$ and $\epsilon_d+U$. At the electron-hole symmetry point ($\epsilon_d=-U/2$) and resonance energies (i.e., for $\epsilon_d=0$ and $\epsilon_d+U=0$), the electron and hole charge current compensate each other, and thus no thermopower is observed at these points (indicated by dashed lines). If the QD energy level lies above the Fermi resonance energy, i.e., $\epsilon_d>0$, more electrons tunnel from the hot reservoir to the cold reservoir. This net flow of electrons from the left to right reservoir gives rise to the negative thermopower (due to negative electron charge $e$) under open circuit condition. However, if $-U/2<\epsilon_d<0$, then the effective level at $\epsilon_d$ lies close to the Fermi resonance energy from below, and more electrons flow from the right to left reservoir, and thermopower becomes positive. The magnitude of thermopower for $-U/2<\epsilon_d<0$ decreases with increasing thermal gradient $\Delta{T}$ as a result of the reduced Coulomb blockade effect due to enhancement in the tunneling of thermally excited electrons from left to the right reservoir. Similarly, one can explain the origin of negative and positive thermopower for $\epsilon_d<-U/2$. The $S$ versus $\epsilon_d$ curves for $\epsilon_d>0$ and $\epsilon_d<-U$ become flat with increasing $\Delta{T}$ due to thermal broadening of the Fermi-Dirac distribution function.\\
The plots for electronic contribution to the thermal conductance $K$ are shown in fig.\hyperref[fig:3]{3}(d-f). It is clear from these plots that $K$ shows minima at the resonance energies ($\epsilon_d=0$ and $\epsilon_d+U=0$) for all values of finite $\Delta{T}$ and maxima at the electron-hole symmetry point ($\epsilon_d=-U/2$) for moderate or relatively larger $\Delta{T}$ (depending on $U$). This behavior differs from the linear response results, which show that at low temperatures, thermal conductance behaves similar to electric conductance, i.e., it shows maxima at the resonance energies and minima at the electron-hole symmetry point. However, at a relatively larger temperature or thermal gradient, the resonant tunneling is significantly reduced, and thus, the electrical conductance, and eventually, $K$ is also suppressed at $\epsilon_d=0$ and $\epsilon_d+U=0$. For $\epsilon_d>0$ and $\epsilon_d<-U$, the thermal conductance is relatively large due to thermally excited electron tunneling and becomes flat as $\Delta{T}$ increases due to the thermal broadening of the Fermi-Dirac distribution function. On the other hand, the electron and hole contributions to thermal conductance add constructively near the electron-hole symmetry point, leading to enhanced $K$ for moderate or relatively larger $\Delta{T}$ if $-U<\epsilon_d<0$. With increasing on-dot Coulomb interaction $U$, the effective QD energy levels $\epsilon_d=-U/2$ and $\epsilon_d+U=U/2$ move further away from resonance and thus require higher $T$ or $\Delta{T}$ for the tunneling of thermally excited electrons.\\
The normalized thermoelectric conversion efficiency at maximum power output $\eta_{P_{max}}$ in fig.\hyperref[fig:3]{3}(g-i) exhibits similar behavior to that of thermopower, i.e., $\eta_{P_{max}}$ becomes minimal at resonances, and the electron-hole symmetry point. The CB effect significantly increases $\eta_{P_{max}}$ between these points but decreases as $\Delta{T}$ and, eventually, $K$ increases further. However, in experimental studies, $\Delta{T}\leq{T}$ and on-dot Coulomb interaction $U$ is the largest energy parameter. Therefore, from now on, we consider these parameter values when calculating $\eta_{P_{max}}$ and other thermoelectric properties. \\
\begin{figure*}[!htb]
\includegraphics
  [width=0.85\hsize]
  {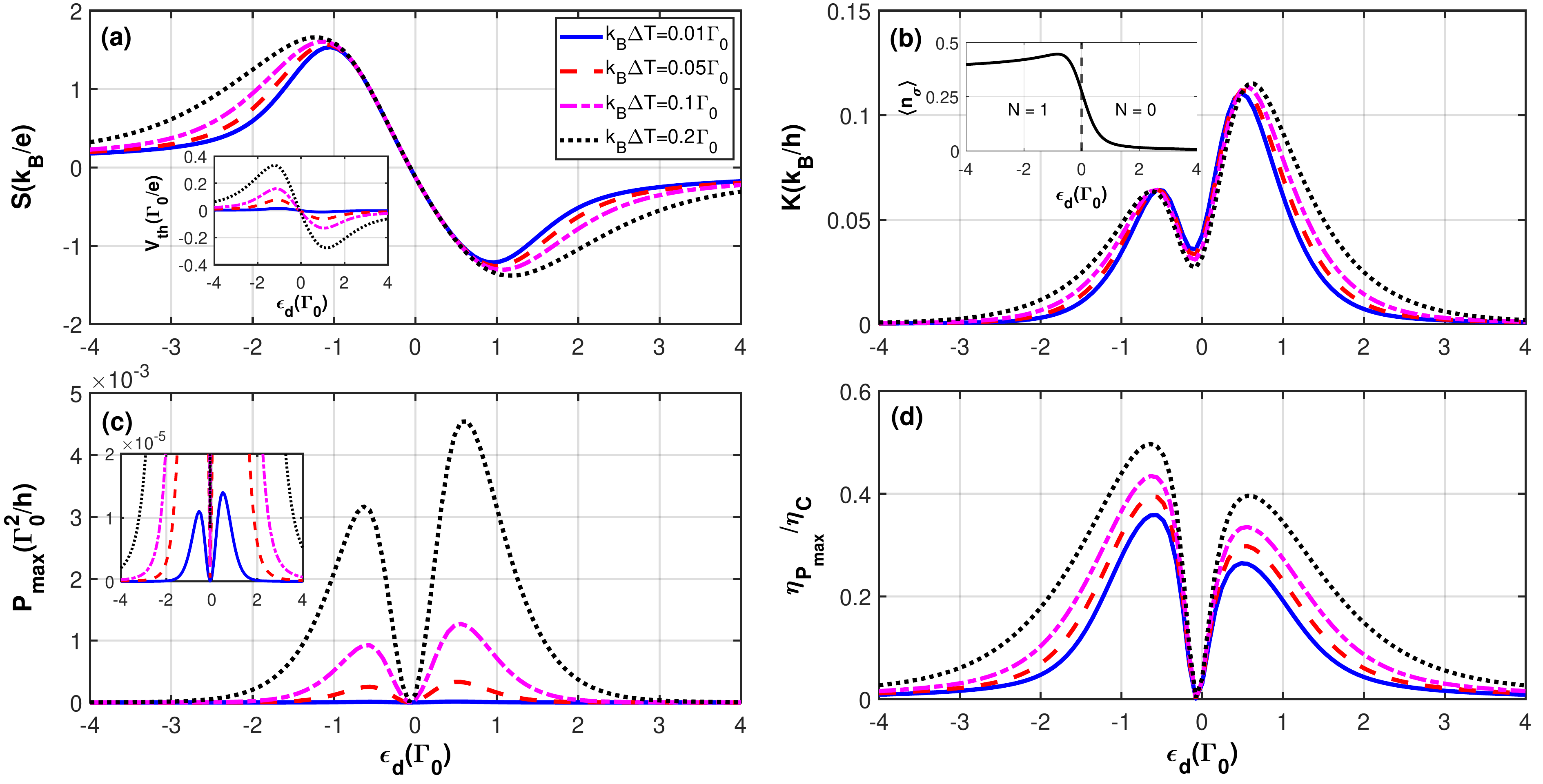}
\caption {Variation of thermopower($S$), thermal conductance($K$), maximum power output($P_{max}$ is maximize with respect to external applied bias), and corresponding normalized efficiency($\eta_{P_{max}}/\eta_C$) with the strongly interacting QD energy level $\epsilon_d$ in the non-linear transport regime for several values of thermal gradient $\Delta T$. The results are obtained within the decoupling scheme beyond Hubbard-\Romannum{1} for $U\rightarrow\infty$ limit. The other parameters are: $k_BT = 0.2\Gamma_0$ and $\Gamma=0.1\Gamma_0$. Inset in (a)  shows the variation of thermovoltage $V_{th}$ with $\epsilon_d$, inset in (b) shows the variation of average QD occupancy $\langle n_{\sigma}\rangle$ with $\epsilon_d$ and inset in (c) shows the close-up view of $P_{max}$ for very small $\Delta T$.}
\label{fig:4}
\end{figure*}
Next, we take into account very strong Coulomb repulsion on the QD energy level, i.e., infinite-$U$ limit. The variation of thermoelectric transport quantities as a function of QD energy level $\epsilon_d$ for different thermal gradients $\Delta{T}$ is shown in fig.\hyperref[fig:4]{4}. For strong on-dot Coulomb repulsion, transport occurs via one single effective energy level $\epsilon_d$. As a result, electrons tunnel sequentially one at a time, and the number of electrons $N$ on the QD is either 1 or 0 depending on the position $\epsilon_d$ w.r.t Fermi resonance energy (see inset in fig.\hyperref[fig:4]{4}(b)). The thermoelectric transport quantities are minimal at the resonance energy $\epsilon_d=0$, as in the non-interacting QD case. However, the peaks in these thermoelectric transport quantities across $\epsilon_d$ are asymmetric due to asymmetry in the tunneling amplitude for even ($N=0$) and odd ($N=1$) electrons on the strongly interacting QD energy level. When $\epsilon_d>0$, the average number of electrons on QD is zero ($N=0$). As a result, electrons with both up and down spin can tunnel to the QD from the reservoir. For $\epsilon_d<0$, the QD energy level is already occupied by a single electron ($N=1$), and only an electron with opposite spin can tunnel to the $QD$ from the reservoir, which leads to a relatively smaller tunneling probability and current. Thus, due to this reason, the peaks in thermal conductance $K$ and maximum power output $P_{max}$ are low when $\epsilon_d<0$ and relatively high when $\epsilon_d>0$. On the other hand, low thermal conductance and large power output are required for optimal $S$ and $\eta_{P_{max}}$. Due to the significant asymmetry in $K$ relative to $P_{max}$, the peaks in $S$ and $\eta_{P_{max}}$ are relatively high when $\epsilon_d<0$. Furthermore, for $\Delta{T}\leq{T}$, thermopower $S$ is independent of $\Delta{T}$ in the range $-\Gamma_0<\epsilon_d<\Gamma_0$ due to linear increase in thermovoltage $V_{th}$(inset in fig.\hyperref[fig:4]{4}(a)). Also, the magnitude of thermal conductance $K$ in fig.\hyperref[fig:4]{4}(b) is not very sensitive to the variation in $\Delta{T}$. The maximum power output $P_{max}$ and normalized efficiency $\eta_{P_{max}}$ are increased by approximately $300$ times and $14\%$ respectively due to a significant increase in thermovoltage and thermocurrent when $k_B\Delta{T}$ is increased from $0.01\Gamma_0$ to $0.2\Gamma_0$.
\subsubsection{Effect of external load resistance ($R$) on the performance of QD heat engine}
\begin{figure*}[!htb]
\includegraphics
  [width=0.85\hsize]
  {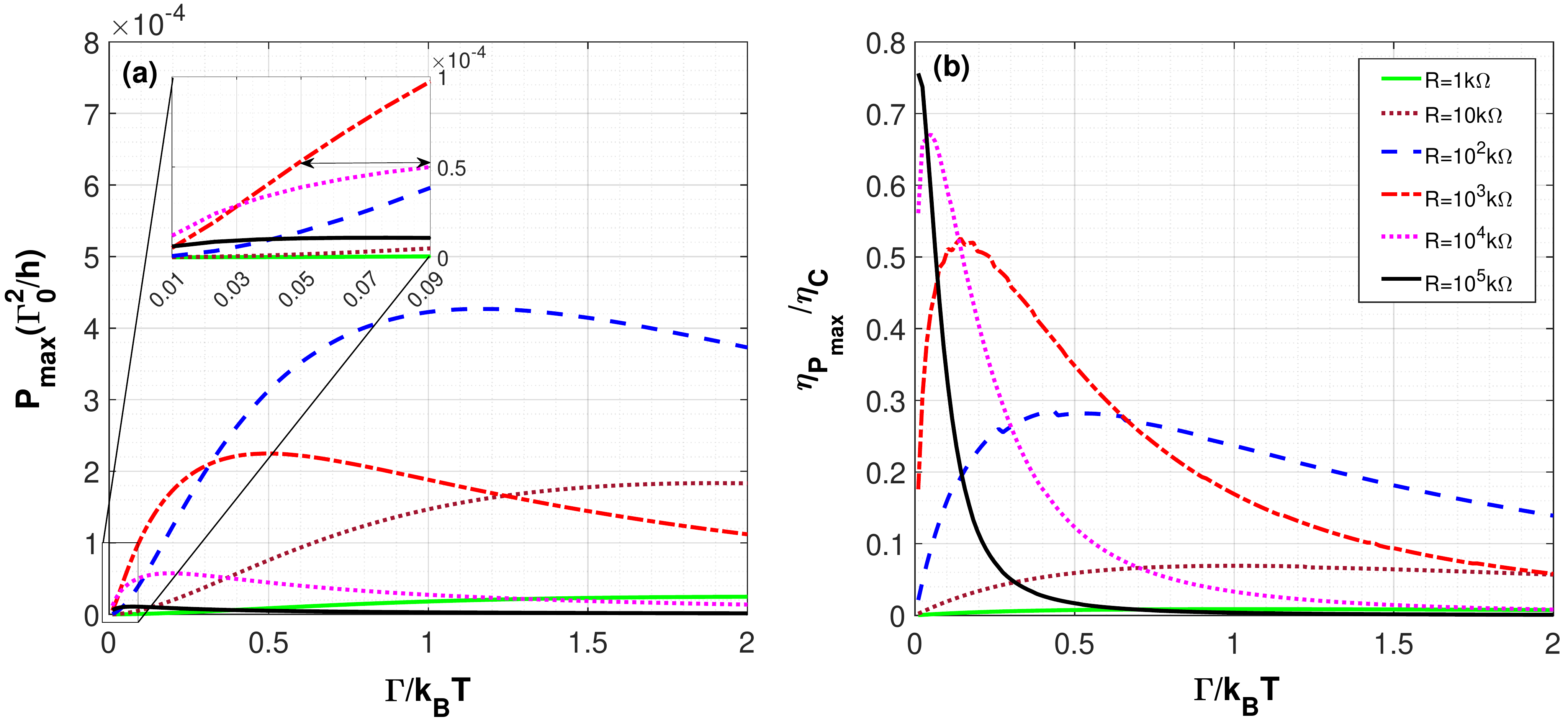}
\caption {Variation of (a) maximum power output $P_{max}$ and (b) corresponding normalized efficiency $\eta_{P_{max}}$ of a strongly interacting (i.e., $U\rightarrow\infty$) single QD particle-exchange heat engine as a function of tunnelling rate ($\Gamma$) for several values of external load resistance $R$. Both $P_{max}$ and $\eta_{P_{max}}$ are first optimized with respect to the QD energy level ($\epsilon_d$). Inset in (a) shows the close-up view of $P_{max}$ for $\Gamma<<k_BT$. The narrow region indicated by a double arrow shows the range of $\Gamma/k_BT$ considered in the experimental study \cite{Josefsson2018}. The other parameters are: $k_BT=0.1\Gamma_0$ and $k_B\Delta T=0.05\Gamma_0$.}
\label{fig:5}
\end{figure*}

In the previous subsection, the thermoelectric performance of a QD heat engine was calculated using an open circuit condition, i.e., an external reverse bias voltage $V_{ext}=(\mu_R-\mu_L)/e$ is applied to counteract the thermally induced charge current. Let us now investigate the effect of external load resistance $R$ on the thermoelectric performance of the single QD heat engine by replacing the external voltage source with a serial load resistance $R$. For finite $R$, the current must self-consistently satisfy Eq.\hyperref[eq:8]{(8)}. In fig.\hyperref[fig:5]{5}, maximum power output $P_{max}$ and corresponding efficiency $\eta_{P_{max}}$ are plotted as a function of $\Gamma/k_BT$ for various values of $R$. These thermoelectric quantities are initially optimized by varying the QD energy level above Fermi energy for each $\Gamma/k_BT$ value. When the dot-reservoir tunneling rate $\Gamma$ is increased, the transmission function becomes broadened, and more electron flows from the hot reservoir to the cold reservoir, generating large thermovoltage. The magnitude of thermovoltage also depends on external serial load $R$ because any voltage drop across $R$ again acts back on the QD. This can be easily understood in two limiting cases, i.e., for short circuit ($R\approx0$) and open circuit ($R\approx\infty$). For a short circuit, there is no voltage drop in the external circuit, i.e., the thermally generated charge current across the QD is instantaneously compensated by the current flowing in the external circuit. Thus no net thermovoltage is generated. On the other hand, for an open circuit ($I_C=0$), the voltage drop is maximum, which causes large thermovoltage. However, power output vanishes in both limits (see fig.\hyperref[fig:5]{5}(a)). Thus, finding the external load $R$ which gives optimal power output at a finite thermoelectric efficiency is necessary for practical applications.
\\
The broadening of transmission function with increasing $\Gamma$ allows more electrons to contribute to power generation. As a result, as illustrated in fig.\hyperref[fig:5]{5}(a), $P_{max}$  increases until the current caused by the temperature difference between the left and right reservoir becomes saturated and reaches a peak value determined by $R$. If the width of the transmission function is further broadened, electrons may start flowing from the right to the left reservoir, which would result in a decrease in $P_{max}$. The shift in the peak value of $P_{max}$ towards a weak dot-reservoir tunneling rate or narrow transmission function with increasing external load $R$ is a manifestation of the maximum power transfer theorem which states that the power transferred from a QD to an external load R is maximal when $R=R_i$ (where $R_i$ is the internal resistance of QD heat engine). For a narrow Dirac delta-like transmission function, $R_i$ is high and requires a large external load to produce the maximum power output. In the present case, the optimal power output can be achieved at either $R\approx{10^{3}k\Omega}$ or $R\approx{10^{2}k\Omega}$ depending on the $\Gamma/k_BT$. On the other hand, as demonstrated in earlier studies, for a Dirac delta-like narrow transmission function, $\eta_{P_{max}}$ approaches the Carnot efficiency, as shown in fig.\hyperref[fig:5]{5}(b)). Due to the narrow energy range with an open circuit ($R\approx\infty$), thermal conductance vanishes faster than electrical conductance, which violates the Widemann-Franz law and causes significant enhancement in $\eta_{P_{max}}$. For a transmission function with finite width, thermal conductance increases rapidly, and the power output cannot compensate for the heat loss, which results in a relatively low thermoelectric conversion efficiency. For example when $R=10^{5}k\Omega$, the value of $\eta_{P_{max}}$ drops by $40\%$ of Carnot efficiency as $\Gamma$ is increased from $0.01k_BT$ to $0.1k_BT$. Further, when the external load resistance $R=10^{5}k\Omega$ is replaced by $R\leq10^{2}k\Omega$ the efficiency $\eta_{P_{max}}$ for $\Gamma<<k_BT$ (say at $\Gamma=0.01k_BT$) significantly reduced from $0.75\eta_C$ to less then $0.1\eta_C$. This drastic reduction in $\eta_{P_{max}}$ is due to vanishing power output or sharp decrease in power output as compare to the thermal conductance for $R\leq10^{2}k\Omega$. Also, as the magnitude of external load resistance is reduced, the peak value of $\eta_{P_{max}}$ shifts towards finite $\Gamma$, similar to $P_{max}$. For $R=10^{2}k\Omega$, optimal power output with a finite efficiency ($\eta_{P_{max}}\approx{0.25\eta_C}$) can be achieved at $\Gamma\approx{k_BT}$.\\
\begin{figure*}[!htb]
\centering
\includegraphics
  [width=0.6\hsize]
  {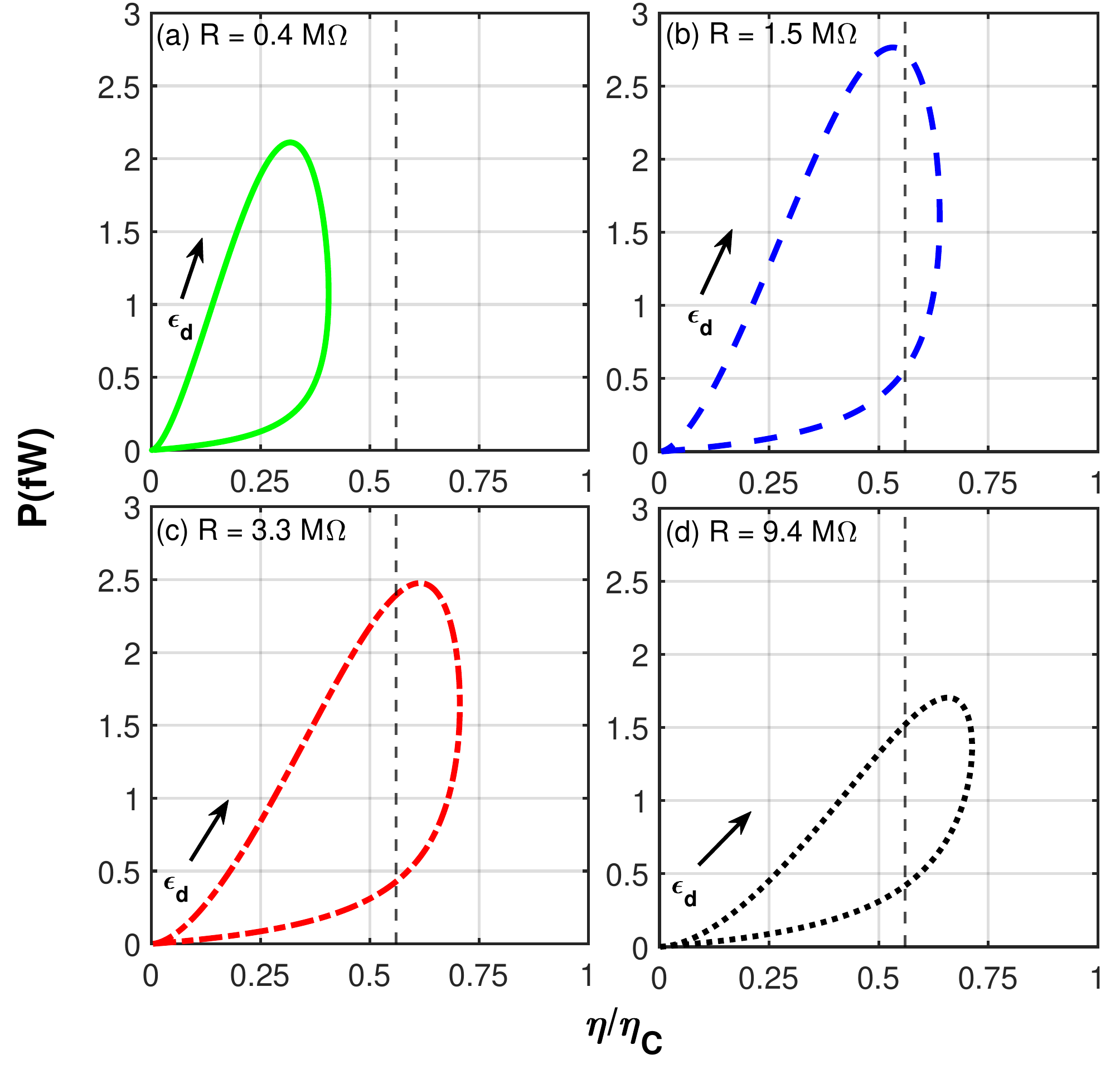}
\caption {The variation of Power output ($P$) versus normalized conversion efficiency ($\eta$) for several values of external load resistance $R$ when QD energy level $\epsilon_d$ or gate voltage $V_G$ is varied (indicated by black arrows). The results are obtained within the decoupling scheme beyond Hubbard-\Romannum{1} for $U\rightarrow\infty$ limit by using Green's function EOM technique. Black dashed lines at $\eta=0.56\eta_C$ indicate the Curzon-Ahlborn efficienct ($\eta_{CA}$) i.e  efficiency corresponding to the maximum power output (in present case at $R=1.5 M\Omega$). The values of the other parameters are similar as considered in the experiment\cite{Josefsson2018} i.e. $T=0.9K$ ($k_BT=0.078 meV$), $\Delta T=0.6K$ ($k_BT=0.052 meV$), $\Gamma=5.8\mu eV$, and $\epsilon_d$ varies from $0\ meV$ to $1\ meV$.}
\label{fig:6}
\end{figure*}
Finally, we compare our infinite-$U$ results with the experiment and real-time diagrammatic (RTD) theory results from Ref.\cite{Josefsson2018}. Since $U$ is the largest energy parameter in the experimental device, thus the infinite-$U$ limit is a valid approach. Further, the dot-reservoir tunneling rate is very low, i.e., $\Gamma<<k_BT$. Thus, experimental results are limited to a narrow region (i.e., $0.05\leq\Gamma/k_BT\leq0.09$), as indicated by a double arrow in the inset of fig.\hyperref[fig:5]{5}(a). Fig.\hyperref[fig:6]{6} shows that the power output $P$ and corresponding thermoelectric conversion efficiency $\eta$ can be optimized for any given load $R$ by varying the QD energy level $\epsilon_d$ (applying gate voltage $V_G$). However, for $\Gamma<<k_BT$, there is always a finite tradeoff between the maximum values of $P$ and $\eta$ because these quantities are not maximized by the same $\epsilon_d$ and $R$. The power output $P$ is maximum for $R=1.5 M\Omega$ with thermoelectric conversion efficiency approximately equal to the  Curzon-Ahlborn efficiency, i.e., $\eta\approx\eta_{CA}=0.56\eta_C$. On the other hand, for $R=3.3 M\Omega$, and $9.4 M\Omega$, the efficiency $\eta$ reaches its maximum value, which is approximately equal to $70\%$ of the Carnot efficiency, i.e., $\eta_{max}=0.7\eta_C$.\\
\begin{figure*}[!htb]
\centering
\includegraphics
  [width=0.75\hsize]
  {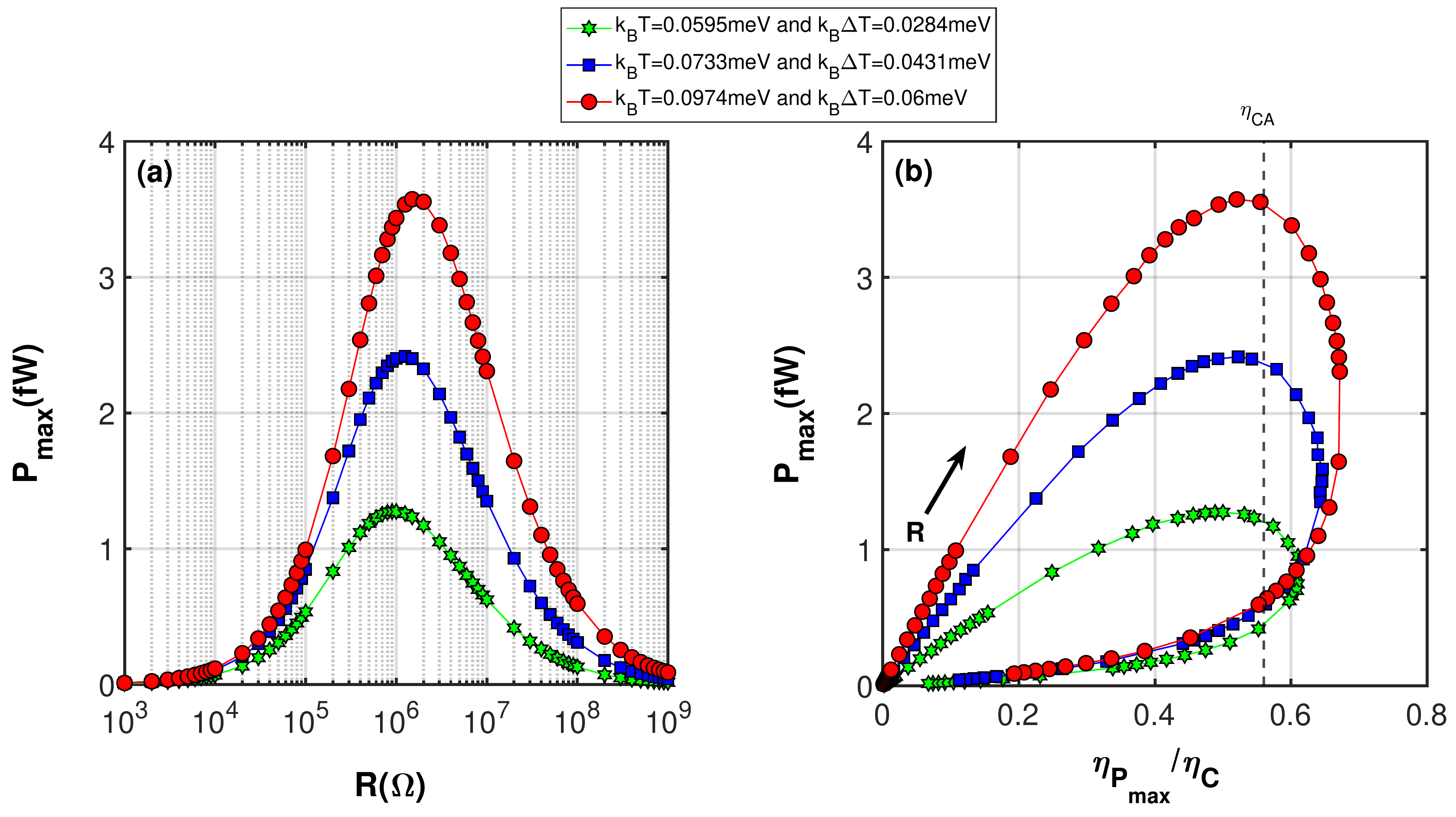}
\caption {The theoretical variation of (a) $P_{max}$ (maximize with respect to $\epsilon_d$) as a function of external load resistance $R$ (in $\Omega$) and (b) $P_{max}$ versus normalized efficiency at maximum power output $\eta_{P_{max}}$ when the external load $R$ is varied for three different set of background temperature and thermal gradient between the reservoirs[i.e. $T=0.69K$, $\Delta T=0.33K$ (\protect\markerone), $T=0.85K$, $\Delta T=0.5K$ (\protect\markertwo) and  $T=1.13K$, $\Delta T=0.7K$ (\protect\markerthree)]. The arrow indicates the direction for increasing $R$. The results are obtained within the decoupling scheme beyond Hubbard-\Romannum{1} for $U\rightarrow\infty$ limit by using Green's function EOM technique. The other parameters are the same as in fig.\hyperref[fig:6]{6} above.}
\label{fig:7}
\end{figure*}
Fig.\hyperref[fig:7]{7} shows the plot for (a) $P_{max}$ and (b) $\eta_{P_{max}}$ as a function of external load resistance $R$ for different background temperatures $T$ and thermal gradients $\Delta{T}$. First, for each $R$, the maximum power output $P_{max}$ is calculated by varying $\epsilon_d$ from $0$ to $1meV$, as done in fig.\hyperref[fig:6]{6}. The load resistance $R$ is then varied between $1k\Omega$ and $1G\Omega$ to find its optimal value $R_P$ i.e. load resistance corresponding to highest $P_{max}$ value. It is important to note that $R_P$ can be calculated directly from the open circuit condition by using $R_P=(V_{th}/2)/(I_C)_{V{th}/2}$. However, because we are interested in analyzing the effect of $R$ on the performance of the QD heat engine, we calculated $R_P$ by varying $R$. Fig.\hyperref[fig:7]{7}(a) shows that $P_{max}$ peaked for $R$ between $1.0$ to $2.0$ $M \Omega$ depending on $T$ and $\Delta{T}$. The peak value of $P_{max}$ nearly doubles as ($T$,$\Delta{T}$) increases from ($0.69K$,$0.33K$) to ($1.13K$,$0.7K$). The optimal load $R_P$ also depends on the ratio $\Gamma/k_BT$, and as previously noted in fig.\hyperref[fig:5]{5}, $R_P$ is close to $0.1M\Omega$ for $\Gamma/k_BT\gtrsim{0.3}$. Fig.\hyperref[fig:7]{7}(b) shows that thermoelectric efficiency at maximum power output $\eta_{P_{max}}$ is peaked for $R$ between $4$ to $10$ $M \Omega$ depending on $T$ and $\Delta{T}$. As ($T$,$\Delta{T}$) increases from ($0.69K$,$0.33K$) to ($1.13K$,$0.7K$), the peak value of $\eta_{P_{max}}$ enhanced by $6\%$ of Carnot efficiency. Further, for $R\approx{10M}\Omega$, the efficiency of strongly interacting single QD heat engine can reach near $70\%$ of Carnot efficiency with a finite power output $P_{max}\approx{2fW}$ and also $\eta_{P{max}}\approx\eta_{CA}$ at the optimal load resistance $R_P$. Our theoretical estimation for the load resistance necessary to achieve the optimal power output and corresponding thermoelectric efficiency are in good agreement with the recent experimental and the RTD theory results.
\section{Conclusion}\label{sec: fourth}
 In summary, we have analyzed a strongly correlated single quantum dot-based thermoelectric particle-exchange heat engine using Keldysh's non-equilibrium Green's function formalism for various order of decoupling schemes in the equation of motion technique. It is shown that the Hartree-Fock mean-field approximation is too simple to correctly describe the Coulomb blockade regime of the QD heat engine. Therefore, we employed Hubbard-\Romannum{1} approximation for a finite on-dot Coulomb interaction and beyond Hubbard-\Romannum{1} for strongly interacting QD, i.e., $U\rightarrow\infty$ limit, to study the thermoelectric transport properties in the Coulomb blockade regime. For finite-$U$, the thermopower and thermoelectric efficiency are significantly enhanced between the electron-hole symmetry and resonance points at a relatively small thermal gradient. When the Coulomb interaction on the QD is very strong, i.e., infinite-$U$, the thermoelectric quantities become asymmetric due to asymmetry in the tunneling amplitude for odd and even electrons, and the maximum power output and corresponding efficiency increases significantly with thermal gradient. For practical applications, however, it is essential to understand the relationship between power output and corresponding efficiency in the presence of finite external load. Therefore, using the infinite-$U$ limit, the effect of dot-reservoir tunneling rate and external load resistance on the optimization of the power output and corresponding thermoelectric efficiency are investigated. It has been demonstrated that $P_{max}$ and $\eta_{P_{max}}$ are peaked at different external load resistance, which differ by approximately one order of magnitude. Due to this finite tradeoff, it is essential to make an adjustment between $P_{max}$ and $\eta_{P_{max}}$ to achieve the best performance in practical applications. These results for the infinite-$U$ limit are in good agreement with recent experimental data and real-time diagrammatic theory results. Thus Green's function EOM technique which is a computationally inexpensive and straightforward analytical method gives reliable results in the Coulomb blockade regime. The present analysis can be extended to examine the optimal performance of other realistic low-dimensional heat engines based on multiple quantum dots and multiple reservoirs within the strong Coulomb blockade regime.
\begin{acknowledgements}
Sachin Verma is presently a research scholar at the department of physics IIT Roorkee and is highly thankful to the Ministry of Education (MoE), India, for providing financial support in the form of a Ph.D. fellowship. The authors also acknowledge the financial support from the DST-SER-1644-PHY 2021-22 research project.
\end{acknowledgements}
\appendix
\section{Hubbard-{\romannum{1}} approximation for finite on-dot Coulomb intearction $U$} \label{Appendix: A}
To solve the single impurity Anderson model Hamiltonian, we apply the Green's function equation of motion (EOM) method in Fourier space with Zubarev notation\cite{Zubarev1960}. The Fourier transform
of the single particle retarded Green’s function of QD ($G_{11\sigma}^r(\omega)=\langle\langle{d_{\sigma}|d^\dagger_{\sigma}}\rangle\rangle$) with spin $\sigma$ must satisfies the following EOM. 
\begin{equation} \label{eq:A1}
\omega\langle\langle{d_{\sigma}|d^\dagger_{\sigma}}\rangle\rangle=\langle\{d_{\sigma},d^\dagger_{\sigma}\}\rangle+\langle\langle[d_{\sigma},\hat{H}]|d^\dagger_{\sigma}\rangle\rangle
\end{equation}
By evaluating commutator and anti-commutator
brackets we drive the following equation for the single particle retarded Green’s function for a correlated QD coupled to metallic source and drain reservoir.
\begin{equation} \label{eq:A2}
\left(\omega-\epsilon_{d}-\sum_{k\alpha}\frac{|V_{\alpha}|^2}{(\omega-\epsilon_{k,\alpha})}\right)\langle\langle{d_{\sigma}|d^\dagger_{\sigma}}\rangle\rangle=1+U\langle\langle{d_{\sigma}n_{\bar{\sigma}}|d^\dagger_{\sigma}}\rangle\rangle                                              
\end{equation}
For a non-zero Coulomb interaction an EOM for a given Green’s function involves higher-order coupled Green’s functions or correlation functions, thus creating a hierarchy of equations. In order to truncate the hierarchy of equations one need a decoupling scheme for higher order correlation functions.\\
In this appendix, we use Hartre-Fock (HF) mean-field and Hubbard-\Romannum{1} decoupling scheme to study the CB effects\cite{Anderson1961,Hubbard1963,Hewson1993}. For simplification we assume that the coupling strength is $k$ independent i.e. $V_{k,\alpha}=V_{\alpha}$ for $V_{k,\alpha}<<D$ (wide band) where $-D\leq\epsilon_{k,\alpha}\leq D$ and the tunneling rate from dot to the $\alpha$-leads is defined by $\Gamma_{\alpha}=2\pi|V_{\alpha}|^2\rho_{0\alpha}$. Where density of states of metallic reservoir ($\rho_{0\alpha}$) is constant in the range of energy around Fermi level (flat band).\\
The HF mean-field decoupling is the lowest order approximation to study the CB effects\cite{Anderson1961}. Within this approximation Eq.\hyperref[eq:A2]{(A2)} is simplified by using following decoupling,
\begin{equation*}
\langle\langle{d_{\sigma}n_{\bar{\sigma}}|d^\dagger_{\sigma}}\rangle\rangle \approx \langle{n_{\bar{\sigma}}}\rangle\langle\langle{d_{\sigma}|d^\dagger_{\sigma}}\rangle\rangle
\end{equation*}
where $\langle{n_{\bar{\sigma}}}\rangle$ denotes the quantum statistical average value of occupation number with spin $\bar{\sigma}$. This decoupling amounts to simply replacing the term $\hat{n}_{d\uparrow}\hat{n}_{d\downarrow}$ in Hamiltonian by effective one-body term $\langle{n}_{d\uparrow}\rangle\hat{n}_{d\downarrow}+\langle{n}_{d\downarrow}\rangle\hat{n}_{d\uparrow}$. Thus, within the HF mean-field approximation the single particle retarded Green's function of electron with spin $\sigma$ on the quantum dot is given by,
\begin{equation} \label{eq:A3}
G_{11\sigma}^{r}(\omega)=\langle\langle{d_{\sigma}|d_{\sigma}^{\dagger}}\rangle\rangle=\left[\omega-\epsilon_d-U\langle{n_{\bar{\sigma}}}\rangle+\sum_{\alpha}\frac{i\Gamma_{\alpha}}{2}\right]^{-1}
\end{equation}
To go beyond the HF mean field approximation the equation of motion for higher order correlation function in Eq.\hyperref[eq:A2]{(A2)} is given by,
\begin{equation} \label{eq:A4}
\left(\omega-\epsilon_{d}-U\right)\langle\langle{d_{\sigma}n_{\bar{\sigma}}|d^\dagger_{\sigma}}\rangle\rangle                                              =\langle{n_{\bar{\sigma}}}\rangle+\sum_{k,\alpha}V^{\ast}_{\alpha}\langle\langle{c_{k\sigma,\alpha}n_{\bar{\sigma}}|d^\dagger_{\sigma}}\rangle\rangle-\sum_{k,\alpha}V^{\ast}_{\alpha}\langle\langle{c_{k\bar{\sigma},\alpha}d_{\bar{\sigma}}^{\dagger}d_{\sigma}|d^\dagger_{\sigma}}\rangle\rangle+\sum_{k,\alpha}V^{\ast}_{\alpha}\langle\langle{c_{k\bar{\sigma},\alpha}^{\dagger}d_{\sigma}d_{\bar{\sigma}}|d^\dagger_{\sigma}}\rangle\rangle                                                                                        
\end{equation}
The Hubbard-\Romannum{1} approximation is next higher order approximation than mean field approximation\cite{Hubbard1963}. Within the Hubbard-\Romannum{1} approximation following decoupling scheme is used in Eq.\hyperref[eq:A4]{(A4)},
\begin{equation*}
\langle\langle{c_{k\sigma,\alpha}{n_{\bar{\sigma}}}|d_{\sigma}^{\dagger}}\rangle\rangle \approx \langle{n_{\bar{\sigma}}}\rangle \langle\langle{c_{k\sigma,\alpha}|d_{\sigma}^{\dagger}}\rangle\rangle 
\end{equation*}
\begin{equation*}
\langle\langle{c_{k\bar{\sigma},\alpha}d_{\bar{\sigma}}^{\dagger}d_{\sigma}|d^\dagger_{\sigma}}\rangle\rangle \approx \langle{d_{\bar{\sigma}}^{\dagger}d_{\sigma}}\rangle \langle\langle{c_{k\bar{\sigma},\alpha}|d_{\sigma}^{\dagger}}\rangle\rangle
\end{equation*}
\begin{equation*}
\langle\langle{c_{k\bar{\sigma},\alpha}^{\dagger}d_{\sigma}d_{\bar{\sigma}}|d^\dagger_{\sigma}}\rangle\rangle \approx \langle{d_{\sigma}d_{\bar{\sigma}}}\rangle \langle\langle{c_{k\bar{\sigma},\alpha}^{\dagger}|d_{\sigma}^{\dagger}}\rangle\rangle                                              
\end{equation*}
We have considered the average $\langle{d_{\sigma}d_{\bar{\sigma}}}\rangle$ as zero because such kind of terms are non-zero only for superconductor. Further, the spin flip processes on the QD are neglected (i.e. $\langle{d_{\bar{\sigma}}^{\dagger}d_{\sigma}}\rangle=0$) and thus the approximation is relevant for temperatures higher than the temperature associated with the Kondo effect($T>>T_K$).\\
Solving coupled Eqs.\hyperref[eq:A2]{(A2)} and \hyperref[eq:A4]{(A4)} using the above Hubbard-\Romannum{1} decoupling gives the single particle retarded Green's function of electron on the QD with spin $\sigma$.
\begin{equation} \label{eq:A5}
G_{11\sigma}^{r}(\omega)=\langle\langle{d_{\sigma}|d_{\sigma}^{\dagger}}\rangle\rangle=\frac{(\omega-\epsilon_d-U(1-\langle{n_{\bar{\sigma}}}\rangle))}{(\omega-\epsilon_d)(\omega-\epsilon_d-U)+\sum_{\alpha}\frac{i\Gamma_{\alpha}}{2}(\omega-\epsilon_d-U(1-\langle{n_{\bar{\sigma}}}\rangle))}
\end{equation}
\section{Beyond Hubbard-{\romannum{1}} for $U\rightarrow\infty$ limit}\label{Appendix: B}
In this appendix, we use a decoupling scheme beyond the Hubbard-\Romannum{1} approximation to study very strong Coulomb replusion on the QD energy level. To go beyond the Hubbard-\Romannum{1} approximation the equation of motion for higher order correlation functions in Eq.\hyperref[eq:A4]{(A4)} is given by
\begin{align} \label{eq:B1}
\begin{aligned}
\left(\omega-\epsilon_{k,\alpha}\right)\langle\langle{c_{k\sigma,\alpha}n_{\bar{\sigma}}|d^\dagger_{\sigma}}\rangle\rangle = V_{\alpha}\langle\langle{d_{\sigma}n_{\bar{\sigma}}|d^\dagger_{\sigma}}\rangle\rangle+
\sum_{k^{'}\alpha}V_{\alpha}^{\ast}\langle\langle{c_{k^{'}\bar{\sigma},\alpha}^{\dagger}c_{k\sigma,\alpha}d_{\bar{\sigma}}|d_{\sigma}^{\dagger}}\rangle\rangle-\sum_{k^{'}\alpha}V_{\alpha}\langle\langle{c_{k\sigma,\alpha}c_{k^{'}\bar{\sigma},\alpha}d_{\bar{\sigma}}^{\dagger}|d_{\sigma}^{\dagger}}\rangle\rangle
\end{aligned}
\end{align}
\begin{align} \label{eq:B2}
\begin{aligned}
\left(\omega-\epsilon_{k,\alpha}\right)\langle\langle{c_{k\bar{\sigma},\alpha}d_{\bar{\sigma}}^{\dagger}d_{\sigma}|d^\dagger_{\sigma}}\rangle\rangle = {} & -\langle{d_{\bar{\sigma}}^{\dagger}c_{k\bar{\sigma},\alpha}}\rangle-V_{\alpha}\langle\langle{d_{\sigma}n_{\bar{\sigma}}|d^\dagger_{\sigma}}\rangle\rangle+\sum_{k^{'}\alpha}V_{\alpha}\langle\langle{c_{k^{'}\bar{\sigma},\alpha}^{\dagger}c_{k\bar{\sigma},\alpha}d_{\sigma}|d_{\sigma}^{\dagger}}\rangle\rangle-\sum_{k^{'}\alpha}V_{\alpha}^{\ast}\langle\langle{c_{k\sigma,\alpha}c_{k^{'}\bar{\sigma},\alpha}d_{\bar{\sigma}}^{\dagger}|d_{\sigma}^{\dagger}}\rangle\rangle
\end{aligned}
\end{align}
\begin{align} \label{eq:B3}
\begin{aligned}
\left(\omega+\epsilon_{k,\alpha}-2\epsilon_d-U\right)\langle\langle{c_{k\bar{\sigma},\alpha}^{\dagger}d_{\sigma}d_{\bar{\sigma}}|d^\dagger_{\sigma}}\rangle\rangle = {} & -\langle{c_{k\bar{\sigma},\alpha}^{\dagger}d_{\bar{\sigma}}}\rangle+V_{\alpha}^{\ast}\langle\langle{d_{\sigma}n_{\bar{\sigma}}|d^\dagger_{\sigma}}\rangle\rangle-
\sum_{k^{'}\alpha}V_{\alpha}^{\ast}\langle\langle{c_{k^{'}\bar{\sigma},\alpha}^{\dagger}c_{k\bar{\sigma},\alpha}d_{\sigma}|d_{\sigma}^{\dagger}}\rangle\rangle\\
& +\sum_{k^{'}\alpha}V_{\alpha}^{\ast}\langle\langle{c_{k^{'}\bar{\sigma},\alpha}^{\dagger}c_{k\sigma,\alpha}d_{\bar{\sigma}}|d_{\sigma}^{\dagger}}\rangle\rangle 
\end{aligned}
\end{align}
Again, these EOMs contain higher order correlation functions. Hence, we close the system of equations by using following decoupling scheme\cite{Lacroix1981}.\\
\begin{equation*}
\langle\langle{c_{k\sigma,\alpha}c_{k^{'}\bar{\sigma},\alpha}d_{\bar{\sigma}}^{\dagger}|d_{\sigma}^{\dagger}}\rangle\rangle \approx -\langle{d_{\bar{\sigma}}^{\dagger}c_{k^{'}\bar{\sigma},\alpha}}\rangle \langle\langle{c_{k\sigma,\alpha}|d_{\sigma}^{\dagger}}\rangle\rangle 
\end{equation*}
\begin{equation*}
\langle\langle{c_{k^{'}\bar{\sigma},\alpha}^{\dagger}c_{k\sigma,\alpha}d_{\bar{\sigma}}|d_{\sigma}^{\dagger}}\rangle\rangle \approx -\langle{c_{k^{'}\bar{\sigma},\alpha}^{\dagger}d_{\bar{\sigma}}}\rangle \langle\langle{c_{k\sigma,\alpha}|d_{\sigma}^{\dagger}}\rangle\rangle 
\end{equation*}
\begin{equation*}
\langle\langle{c_{k^{'}\bar{\sigma},\alpha}^{\dagger}c_{k\bar{\sigma},\alpha}d_{\sigma}|d_{\sigma}^{\dagger}}\rangle\rangle \approx \langle{c_{k^{'}\bar{\sigma},\alpha}^{\dagger}c_{k\bar{\sigma},\alpha}}\rangle \langle\langle{d_{\sigma}|d_{\sigma}^{\dagger}}\rangle\rangle 
\end{equation*}
The above averages can be further simplified as\cite{Meir1991} \\
\begin{equation}  \label{eq:B4}
\langle{c_{k^{'}{\bar{\sigma}},\alpha}^{\dagger}d_{\bar{\sigma}}}\rangle=\langle{d_{\bar{\sigma}}^{\dagger}c_{k^{'}{\bar{\sigma}},\alpha}}\rangle\approx 0
\ \ \ \& \ \ \ 
\langle{c_{k^{'}{\bar{\sigma}},\alpha}^{\dagger}c_{k\bar{\sigma},\alpha}}\rangle=\langle{c_{k\bar{\sigma},\alpha}^{\dagger}c_{k^{'}{\bar{\sigma}},\alpha}}\rangle\approx \delta_{kk^{'}} f_{\alpha}(\epsilon_{k,\alpha})
\end{equation}
Where $f_{\alpha}(\epsilon_{k,\alpha})$ is the Fermi-Dirac distribution function of $\alpha$ reservoir($\alpha\in L,R$).\\
Above assumption in Eq.\hyperref[eq:B4]{(B4)} is valid approach for describing cotunneling events for strong on-dot Coulomb repulsion at temperature higher then the temperature associated with Kondo effect i.e. $T>>T_K$. After solving five coupled Eqs.\hyperref[eq:A2]{(A2)}, \hyperref[eq:A4]{(A4)}, \hyperref[eq:B1]{(B1)}, \hyperref[eq:B2]{(B2)} and \hyperref[eq:B3]{(B3)} within above decoupling scheme, the expression for the single particle retarded Green's function of electron with spin $\sigma$ on the quantum dot is given by,
\begin{equation} \label{eq:B5}
G_{11\sigma}^{r}(\omega)=\langle\langle{d_{\sigma}|d_{\sigma}^{\dagger}}\rangle\rangle=\left[\cfrac{1+\cfrac{U\langle{n_{\bar{\sigma}}}\rangle}{\omega-\epsilon_d-U-\Sigma_0-\Sigma_2}}{\omega-\epsilon_d-\Sigma_0+\cfrac{U\Sigma_1}{\omega-\epsilon_d-U-\Sigma_0-\Sigma_2}}\right]
\end{equation}
where
\begin{equation}  \label{eq:B6}
\Sigma_0=\sum_{k\alpha}\left|V_{\alpha}\right|^2\left[\frac{1}{\omega-\epsilon_{k,\alpha}}\right]
\ \ \ \& \ \ \ 
\Sigma_i=\sum_{k\alpha}A_{k,\alpha}^{(i)}\left|V_{\alpha}\right|^2\left[\frac{1}{\omega-\epsilon_{k,\alpha}}+\frac{1}{\omega-2\epsilon_d-U+\epsilon_{k,\alpha}}\right]
\end{equation}
with $A_{k,\alpha}^{(1)}=f_{\alpha}(\epsilon_{k,\alpha})$, and $A_{k,\alpha}^{(2)}=1$ for $i=1,2$.\\
Expressions in Eq.\hyperref[eq:B6]{(B6)} can be simplify further by replacing $\sum_k\rightarrow\int{\rho(\epsilon_{k,\alpha})\,d\epsilon_{k,\alpha}}$ and then solving these equations analytically using the complex integration for the flat ($\rho(\epsilon_{k,\alpha})\rightarrow\rho_{0\alpha}$) and wide band limit ($D\rightarrow\infty$).\\
For strong Coulomb blockade limit (i.e. $U\rightarrow\infty$) above Green's function (Eq.\hyperref[eq:B5]{(B5)}) become,
\begin{equation} \label{eq:B7}
G_{11\sigma}^{r, U\rightarrow\infty}(\omega)=\cfrac{1-\langle n_{\bar{\sigma}}\rangle}{\omega-\epsilon_d-\Sigma_0-\Sigma_1}
\end{equation}\\
with \ \ 
$
\Sigma_0=-\sum_{\alpha}\cfrac{i\Gamma_{\alpha}}{2}
$,
\ \ and \ \ 
$
\Sigma_1=-\sum_{\alpha}\cfrac{\Gamma_{\alpha}}{2\pi}\left[\psi\left({\cfrac{1}{2}+\cfrac{\omega}{2\pi i k_BT_{\alpha}}}\right)\right]
$\\
Where $\psi(...)$ is the digamma function.

\end{document}